\documentclass[twocolumn,           
               showpacs,            
               preprintnumbers,     
               aps,                 
               prd,                 
               a4paper,             
               superscriptaddress,  
               unsortedaddress,     
               nofootinbib,           
               tightenlines,        
               floats               
               ]{revtex4}

\usepackage{graphicx}
\usepackage{latexsym}
\usepackage{amsmath,amssymb}        
\usepackage[draft=false]{hyperref}

\newcommand{\Sc}{{\rm S}}

\newcommand{\beq}{\begin{equation}}
\newcommand{\beqn}{\begin{eqnarray}} 
\newcommand{\eeq}{\end{equation}}
\newcommand{\eeqn}{\end{eqnarray}}
\newcommand{\beqa}{\begin{eqnarray}}
\newcommand{\eeqa}{\end{eqnarray}}
\newcommand{\bea}{\begin{eqnarray}}
\newcommand{\eea}{\end{eqnarray}}

\def\beq{\begin{equation}}
\def\eeq{\end{equation}}

\begin{document}

\title{Constraints on braneworld inflation from CMB anisotropies}
\author{Shinji Tsujikawa}
\affiliation{Institute of Cosmology and Gravitation,
University of Portsmouth, Portsmouth PO1~2EG, \\
United Kingdom}
\author{Andrew R. Liddle}
\affiliation{Astronomy Centre, University of Sussex, 
             Brighton BN1 9QH, United Kingdom}
\date{\today} 
\pacs{98.80.Cq \hfill astro-ph/0312162}
\preprint{astro-ph/0312162}

\begin{abstract}
We obtain observational constraints on Randall--Sundrum type II braneworld
inflation using a compilation of data including WMAP, the
2dF and latest SDSS
galaxy redshift surveys. We place constraints on three classes of 
inflation models (large-field, small-field and hybrid models) 
in the high-energy regime, which exhibit different behaviour 
compared to the low-energy case. The quartic potential 
is outside the $2\sigma$ observational contour bound for a number of
$e$-folds less than 60, and steep inflation driven by an exponential 
potential is excluded because of its high tensor-to-scalar ratio.
It is more difficult to strongly constrain small-field and hybrid models
due to additional freedoms associated with the potentials, but we obtain upper 
bounds for the energy scale of inflation and the model parameters
in certain cases. We also discuss possible ways to break 
the degeneracy of consistency relations and inflationary observables.  
\end{abstract}

\maketitle

\section{Introduction}                            

The recent publication of data from the Wilkinson Microwave Anisotropy Probe 
(WMAP) \cite{Spergel:2003cb} has brought the global cosmological 
dataset to a precision where it 
seriously constrains inflationary models 
\cite{Peiris:2003ff,Barger:2003ym,Bridle,Kinney:2003uw,Leach:2003us,Tegmark}. 
The observations show strong support 
for the standard inflationary predictions of a flat Universe with adiabatic 
density perturbations, and in particular, the viable parameter space of 
slow-roll inflation models in the standard cosmology has been significantly 
narrowed. We are now entering a golden age where the physics in the 
early universe can be probed by upcoming high-precision
observational data.

It is now possible also to impose observational constraints on inflation in 
non-standard cosmologies, the archetypal example being the braneworld cosmology, 
and in particular the Randall--Sundrum Type II model (RSII) \cite{RSII} which is 
the one most investigated in the literature. While it has been shown that 
observations of the primordial spectra cannot distinguish between the standard 
cosmology and the braneworld \cite{Liddle01}, specific constraints on the 
potential driving inflation will be different between those scenarios. Liddle 
and Smith \cite{Liddle:2003gw} recently imposed the first constraints on such 
models, studying the case of monomial potentials $V \propto \phi^p$ in the 
RSII braneworld and finding that the constraints on $p$ tighten in the 
braneworld regime.

In this paper we aim to make a much more general analysis of constraints on 
inflationary models in the RSII braneworld, under the assumption that 
inflation takes place in the high-energy regime of such theories. 
We will use the current observational datasets, including the latest Sloan 
Digital Sky Survey (SDSS) power 
spectrum data \cite{SDSS}, 
and seek to impose constraints for a range of different types of 
inflationary model. 

\section{Formalism}                            

In the RSII model \cite{RSII}, 
where matter fields are confined to the brane, 
the Einstein  equations can written 
as \cite{SMS}: 
\beq
^{(4)}G_{\mu\nu}=-\Lambda_4 g_{\mu\nu}+\frac{8\pi}{m_{{\rm Pl}}^2} 
T_{\mu\nu}+ \left(\frac{8\pi}{M_5^3}\right)^2
\pi_{\mu\nu}-E_{\mu\nu},
\label{geq}
\eeq
where $T_{\mu\nu}$ and $\pi_{\mu\nu}$ represent the energy--momentum 
tensor on the brane and a quadratic term in $T_{\mu\nu}$, respectively.  
$E_{\mu\nu}$ is a part of the 5-dimensional Weyl tensor, which carries the 
information about the bulk.  The 4- and 5-dimensional 
Planck scales, $m_{{\rm Pl}}$ and $M_5$, are related via the 
3-brane tension, $\lambda$, as 
\beq
\lambda=\frac{3}{4\pi}\frac{M_5^6}{m_{{\rm Pl}}^2}\,.
\label{M5}
\eeq
Hereafter the 4-dimensional cosmological constant  
$\Lambda_4$ is assumed to be zero. 

Adopting a flat Friedmann--Robertson--Walker (FRW) metric as a 
background spacetime on the brane, the Friedmann 
equation becomes
\beq
H^2 \equiv \left(\frac{\dot{a}}{a}\right)^2=\frac{8\pi}{3m_{{\rm Pl}}^2} 
\rho \left(1+\frac{\rho}{2\lambda}\right)\,,
\label{hubble}
\eeq
where $a$, $H$, and $\rho$ are the scale factor, the Hubble parameter, 
and the energy density of the matter
on the brane, respectively.  We ignored the so-called `dark radiation', 
$E_{\mu\nu}$, which decreases as $\sim a^{-4}$ during 
inflation (we caution that this can be important in considering perturbations at 
later stages of cosmological 
evolution \cite{Koyama:2003be}). 
At high energies the $\rho^2$ term is expected to play an important role in 
determining the evolution of the Universe.  

The inflaton field $\phi$, confined to the brane, satisfies 
the Klein--Gordon equation 
\beq
\ddot{\phi}+3H\dot{\phi}+V'(\phi)=0\,,
\label{phi}
\eeq 
where $V(\phi)$ is the inflaton potential and 
a prime denotes a derivative with respect to $\phi$.
The quadratic contribution in Eq.~(\ref{hubble}) 
increases the Hubble expansion rate during inflation,  
which makes the evolution of the inflaton slower
through Eq.~(\ref{phi}).
Combining Eq.~(\ref{hubble}) with Eq.~(\ref{phi}),
we get the following 
equation \cite{Maartens:1999hf,Tsujikawa:2000hi}
\beqa
\frac{\ddot{a}}{a}=\frac{8\pi}{3m_{{\rm Pl}}^2}
\left[(V-\dot{\phi}^2)+\frac{\dot{\phi}^2+2V}{8\lambda}
(2V-5\dot{\phi}^2)\right]\,.
\label{ddota}
\eeqa
The condition for inflation is $\ddot{a}>0$,
which reduces to the standard expression $V>\dot{\phi}^2$ for 
$(\dot{\phi}^2+2V)/8\lambda \ll 1$.
In the high-energy case, this condition corresponds to 
$2V>5\dot{\phi}^2$, which 
means that inflation ends around
$2V \simeq 5\dot{\phi}_f^2$.
Making use of the slow-roll conditions 
in Eqs.\,(\ref{hubble}) and (\ref{phi}),  
the end of inflation is characterized by 
\beq
\frac{V^3(\phi_f)}{V'^2(\phi_f)} \simeq
\frac{5\lambda m_{{\rm Pl}}^2}{24\pi}\,.
\label{end}
\eeq 

The amplitudes of scalar and tensor perturbations generated in 
RSII inflation are given as \cite{Maartens:1999hf,Langlois2} 
\begin{eqnarray}
\label{scala}
A_{{\rm S}}^2 & = &
\frac{512 \pi}{75 m_{{\rm Pl}}^6} \, \frac{V^3}{V'^2} \, 
\left( 1 + \frac{V}{2\lambda} \right)^3 \Biggr|_{k=aH} \,, \\
A_{{\rm T}}^2 & = & 
\frac{4}{25\pi} \, \frac{H^2}{m_{{\rm Pl}}^2} \, 
F^2(x) \Biggr|_{k=aH}
\label{tensa} \,,
\end{eqnarray}
where $x=Hm_{{\rm Pl}}\sqrt{3/(4\pi\lambda)}$
and 
\begin{eqnarray}
F(x) =\left[\sqrt{1+x^2} - x^2 \sinh^{-1}(1/x) \right]^{-1/2} 
\label{F}\,.
\end{eqnarray}
The right hand sides of Eqs.~(\ref{scala}) and (\ref{tensa}) are evaluated  
at Hubble radius crossing, $k=aH$ (here $k$ is the comoving 
wavenumber).

Defining the spectral indices of scalar and tensor perturbations
as
\begin{equation}
n_{{\rm S}}-1 \equiv \frac{{\rm d} \ln A_{{\rm S}}^2}
{{\rm d} \ln k}\Biggr|_{k=aH}\,,
~~~
n_{{\rm T}} \equiv \frac{{\rm d} \ln A_{{\rm T}}^2}
{{\rm d} \ln k}\Biggr|_{k=aH} \,,
\end{equation}
and making use of the slow-roll conditions
in Eqs.~(\ref{hubble}) and (\ref{phi}),
one finds \cite{Maartens:1999hf,Huey:2001ae}
\begin{eqnarray}
\label{ntiltS}
n_{{\rm S}} - 1 = -6 \epsilon+ 2 \eta 
\quad , \quad 
n_{{\rm T}} =-\frac{2}{N'} \frac{x'}{x}
\frac{F^2}{\sqrt{1+x^2}}\,.
\label{ntiltT}
\end{eqnarray}
where $\epsilon$ and $\eta$ are slow-roll parameters, defined by 
\begin{eqnarray}
\epsilon & \equiv & \frac{m_{{\rm Pl}}^2}{16\pi} \, \left( 
	\frac{V'}{V} \right)^2 \; \frac{1 + V/\lambda}{\left(1 +
	V/2\lambda \right)^2} \,, \\
\eta & \equiv & \frac{m_{{\rm Pl}}^2}{8\pi} \, \frac{V''}{V} \; 
	\frac{1}{1+V/2\lambda}  \,,
\label{eta}
\end{eqnarray}
together with the number of $e$-folds
\begin{equation}
\label{efolds}
N \simeq - \frac{8\pi}{m_{{\rm Pl}}^2} \int^{\phi_f}_{\phi} 
\frac{V}{V'}
\left( 1+\frac{V}{2\lambda} \right) d\phi \,.
\end{equation}
Here $\phi_f$ is the value of the inflaton
at the end of inflation.

We shall define the ratio of tensor to scalar perturbations as
\begin{equation}
R \equiv 16 \, \frac{A_{{\rm T}}^2}{A_{{\rm S}}^2} \,,
\label{R}
\end{equation}
which coincides with the definition of $R$ in 
Refs.~\cite{Peiris:2003ff,Barger:2003ym,Tegmark}
$R\simeq 16\epsilon$, in the low-energy limit.
{}From Eqs.~(\ref{scala}), (\ref{tensa}), (\ref{ntiltT}) and (\ref{R}),
one can show that the following consistency relation holds independent of the 
brane 
tension, $\lambda$, as \cite{Huey:2001ae}
\begin{equation}
n_{\rm T}=-R/8\,.
\label{consistency}
\end{equation}

That the consistency equation is unchanged in the RSII braneworld means that the 
perturbations do not contain any extra information as compared to the standard 
cosmology. In particular, this means that they cannot be used to determine the 
brane tension $\lambda$; for any value of $\lambda$ a potential can always be 
found to generate any observed spectra \cite{Liddle01}. This result has a nice 
expression in terms of the horizon-flow parameters defined by 
\cite{STG,Sam} 
\begin{equation}
\epsilon_0=\frac{H_{\rm inf}}{H}\,,~~~
\epsilon_{i+1}=\frac{{\rm d ln} |\epsilon_i|}{{\rm d}N }\,,~~~
(i \ge 0)\,,
\label{hflow}
\end{equation}
where $H_{\rm inf}$ is the Hubble rate at some chosen time.
Then we have 
\begin{eqnarray}
n_{\Sc} &=& 1-2\epsilon_1-\epsilon_2,~n_{{\rm T}} = -2\epsilon_1,~
R= 16\epsilon_1\,, \\
\alpha_{\Sc}& =& -2\epsilon_1 \epsilon_2-
\epsilon_2 \epsilon_3\,,~
\alpha_{\rm T}=-2\epsilon_1 \epsilon_2 \quad 
({\rm for}~~V/\lambda \ll 1)\nonumber 
\label{lowslow}
\end{eqnarray}
and 
\begin{eqnarray}
\label{highslow}
n_{\Sc} &=& 1-3\epsilon_1-\epsilon_2,~n_{{\rm T}} = -3\epsilon_1,~
R=24\epsilon_1\,, \\
\alpha_{\Sc} &=& -3\epsilon_1 \epsilon_2-
\epsilon_2 \epsilon_3\,,~
\alpha_{\rm T}=-3\epsilon_1 \epsilon_2  
\quad ({\rm for}~~V/\lambda \gg 1)\nonumber 
\end{eqnarray}
where $\alpha_{{\rm S},{\rm T}} \equiv dn_{{\rm S},{\rm T}}/d\ln k$ are the 
runnings of the two spectra. We see that these two sets of expressions become 
identical if one associates $2\epsilon_1$ in the low-energy limit with 
$3\epsilon_1$ in the high-energy limit.

The upshot of this correspondence is that a separate likelihood analysis of 
observational data is not needed for the braneworld scenario, as observations 
can be used to constrain the same parametrization of the spectra produced. 
However, when those constraints are then interpreted in terms on the form of 
the inflationary potential, differences will be seen depending on the regime we 
are in. For the remainder of this paper, we will obtain constraints under the 
assumption that we are in the high-energy regime. Our work extends that of 
Liddle and Smith \cite{Liddle:2003gw} who examined only monomial potentials, 
though they did so for a general $\lambda$.

\section{Likelihood analysis}                            

In order to compare the theoretical predictions of braneworld inflation 
with observed CMB anisotropies, we run the CAMB program developed 
in Ref.~\cite{antony1} coupled to the 
CosmoMc (Cosmological Monte Carlo) code \cite{antony2}.
This code makes use of 
a Markov-chain Monte Carlo method to derive the likelihood
values of model parameters.  
In addition to the data sets from WMAP \cite{WMAP},
we include the band-powers on smaller scales corresponding to 
$800 < l <2000$, from the VSA~\cite{VSA},  CBI~\cite{CBI},
ACBAR~\cite{ACBAR}, and the 2dF  \cite{2dF} 
and latest SDSS galaxy redshift surveys \cite{SDSS}. We include both 2dF and 
SDSS under the assumption that they can be treated as statistically independent, 
but in fact little difference arises if either one is dropped.
The CosmoMc code generates a large set of power spectra for 
given values of cosmological and inflationary model parameters,
and finds the likelihood values of parameters by comparing the
temperature (TT) and temperature--polarization 
cross-correlation (TE) anisotropy spectra and the matter power spectrum with  
recent data.
 
The WMAP team \cite{Peiris:2003ff} carried out the likelihood analysis
by varying the four quantities
$A_{\rm S}$, $R$, $n_{\rm S}$ and $\alpha_{\rm S}$.
The quantities $n_{\rm T}$ and $\alpha_{\rm T}$ are related to those by 
consistency equations, and $\alpha_{\rm T}$ has anyway always been ignored so 
far in parameter fits as its cosmological consequences are too subtle for 
current or near-future data to detect.

In adopting an expansion of the spectra, one should be careful about convergence 
criteria. The power spectrum $A_{\rm S}^2(k)$
is generally expanded in the form \cite{Lidsey:1995np}
\begin{eqnarray}
{\rm ln}\,A_{\rm S}^2(k)&=& {\rm ln}\,A_{\rm S}^2(k_*)
+(n_{\rm S}-1)\,{\rm ln} \left(\frac{k}{k_*}\right) \nonumber \\
& & + \frac{\alpha_{\rm S}}{2}\,{\rm ln}^2\left(\frac{k}{k_*}\right)
+\cdots\,,
\end{eqnarray}
where $k_*$ is some pivot wavenumber.
In order for this Taylor expansion to be valid, 
we require the following condition \cite{Leach:2003us}
\begin{equation}
|n_{\rm S}-1| \gg \left|\frac{\alpha_{\rm S}}{2}
{\rm ln} \left(\frac{k}{k_*}\right)\right|\,.
\end{equation}
For the maximum values $|{\rm ln} \left(k/k_*\right)| \sim 10^{4}$
and $|n_{\rm S}-1| \lesssim 0.07$, one gets the convergence
criterion $|\alpha_{\rm S}| \lesssim 0.03$.
If this condition is not imposed, the likelihood 
results are not expected to be completely reliable.
We numerically found that the ratio $R$ shifts toward larger likelihood values
if the upper bound of $\alpha_{\rm S}$ is chosen to be 
greater than 0.03.
This implies that it is important to choose an appropriate
prior for $\alpha_{\rm S}$ in order to get a good 
convergence for inflationary model parameters.

When one performs a likelihood analysis
using horizon-flow parameters, the $2\sigma$ upper limits
of $\epsilon_1$ and $\epsilon_2$ were found to be 
$0<\epsilon_1 <0.032$ and $-0.15<\epsilon_2<0.08$ 
in the low-energy limit ($V/\lambda \to 0$) \cite{Leach:2003us}.
Since $\epsilon_3$ is poorly constrained and is consistent with 
zero, it could be set to zero and then the running ranges $-0.0096<\alpha_{\rm 
S}<0.0051$ from 
Eq.~(\ref{lowslow}), which is well inside
the convergence criterion.

\begin{figure}
\includegraphics[scale=0.6]{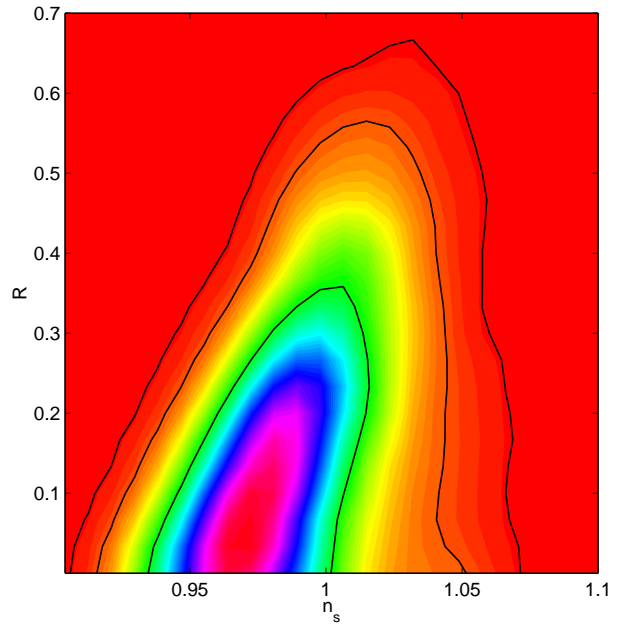} 
\caption{2D posterior constraints in the $n_{\rm S}$-$R$ 
plane.  We also show the $1\sigma$, $2\sigma$ 
and $3\sigma$ contour bounds.
The region with a light colour corresponds to the 
likelihood region.
The ratio $R$ is constrained to be $R<0.57$ 
at the $2\sigma$ level. 
}
\label{nr}
\end{figure}

The high-energy case ($V/\lambda \to \infty$)
corresponds to changing the above $\epsilon_1$ to $(3/2)\epsilon_1$.
Since $n_{\rm T}$, $\alpha_{\rm S}$ and $\alpha_{\rm T}$
are written in terms of $n_{\rm S}$, $R$ and $\epsilon_3$,
from Eqs.~(\ref{consistency}) and (\ref{highslow}), we perform 
the likelihood analysis by varying four quantities: $n_{\rm S}$, $R$, 
$\epsilon_3$ and $A_{\rm S}^2$. 
This is equivalent to varying three horizon-flow parameters
($\epsilon_1, \epsilon_2, \epsilon_3$) in Eq.~(\ref{highslow})
in addition to $A_{\rm S}^2$.
We put a prior $0.8<n_{\rm S}<1.15$ and $0<R<0.7$,
in which case the convergence criterion,
$|\alpha_{\rm S}| \lesssim 0.03$, is satisfied.
We found that $\epsilon_3$ is poorly constrained 
as pointed out in Ref.~\cite{Leach:2003us}, 
which means that 
the present observation does not reach the level
to constrain the higher-order slow-roll parameters. 
Two dimensional observational constraints in terms of 
$n_{\rm S}$ and $R$ are plotted in Fig.~\ref{nr}.
The allowed range of inflationary parameters is 
tighter than the results
by Barger et al.\,\cite{Barger:2003ym},
since we implement several independent cosmological 
parameters in addition to WMAP measurement.
Our results are consistent with the recent work 
by the SDSS group \cite{Tegmark}.

\begin{figure}
\includegraphics[scale=0.47]{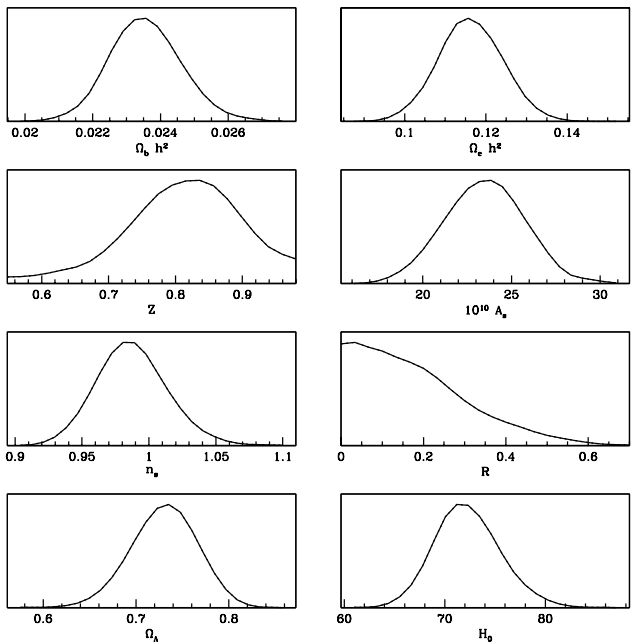} 
\caption{Marginalized probability distributions 
of the cosmological \& inflationary parameters. 
We find that the distribution is 
well described by a Gaussian.}
\label{prob}
\end{figure}

We varied 4 cosmological parameters 
($\Omega_b h^2$, $\Omega_c h^2$, $Z=e^{-2\tau}$, $H_0$)
as well in addition to 4 inflationary variables
by assuming a flat $\Lambda$CDM universe.
Here $\Omega_b h^2$ and $\Omega_c h^2$ are the baryon and
dark matter density, $\tau$ is the optical depth, and 
$H_0$ is the Hubble constant.
As seen in Fig.~\ref{prob}, the likelihood values of these
basic cosmological parameters agree well with past works 
\cite{Peiris:2003ff,Barger:2003ym,Bridle,Kinney:2003uw,Leach:2003us,Tegmark}.

\section{Constraints on braneworld inflation}                            

In this section we shall consider constraints on 
single-field braneworld inflation
in the high-energy case ($V/\lambda \gg 1$).
We can classify models of inflation in the following 
way \cite{Kolb}.
The first class (type I) is the ``large-field" model,
in which the initial value of the inflaton is large 
and it rolls down toward the potential minimum at smaller $\phi$.
Chaotic inflation \cite{Linde83} is one of 
the representative models of this class.
The second class (type II) is the ``small-field" model, in which 
the inflaton field is small initially and slowly evolves toward the potential 
minimum at larger $\phi$.  New inflation \cite{Newinf} and natural 
inflation \cite{Natural} are the examples of this type. 
The third one (type III) is the hybrid (double) inflation model 
\cite{hybrid,hybrid2}, in which inflation ends by a phase transition 
triggered by the presence of the second scalar field (or after a second 
phase of inflation following the phase transition).

When $V/\lambda \gg 1$ we have 
\begin{eqnarray}
\label{highnS}
n_{\rm S}-1 &=& -\frac{m_{{\rm Pl}}^2}{2\pi}
\frac{\lambda}{V} \left[3 \left(\frac{V'}{V}\right)^2
-\frac{V''}{V} \right]\,, \\
R &=& \frac{6m_{{\rm Pl}}^2}{\pi}\left(\frac{V'}{V}\right)^2
\frac{\lambda}{V}\,.
\label{highR}
\end{eqnarray}
In this case the relation between $n_{\rm S}$ and $R$
can be written as
\begin{equation}
R=4(1-n_{\rm S})+8\eta\,.
\end{equation}

The border of large-field and small-field models is given
by the linear potential
\begin{equation}
V=m\phi\,.
\label{linear}
\end{equation}
Since $V''$ vanishes in this case (i.e., $\eta=0$), 
the spectral index of scalar perturbations is 
$n_{\rm S}-1=-6\epsilon$ from Eq.~(\ref{ntiltS}).
In this case we have 
\begin{eqnarray}
R= 4(1-n_{\rm S})\,.
\label{nandR}
\end{eqnarray}
%

\begin{figure}
\includegraphics[scale=0.6]{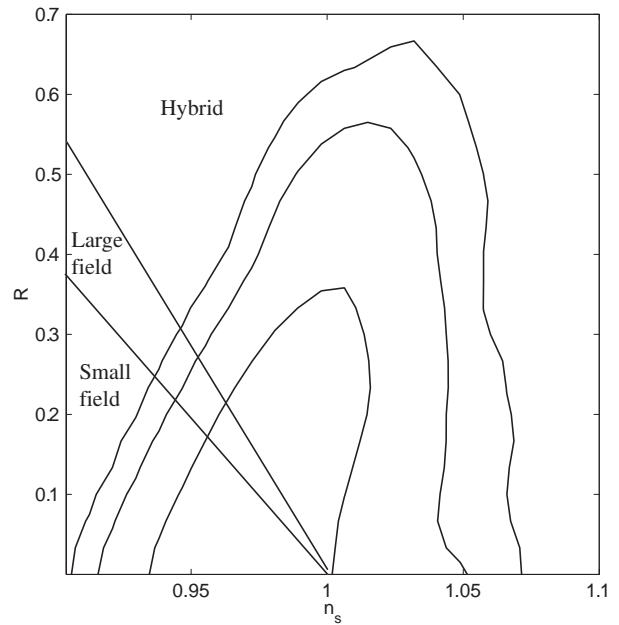} 
\caption{Classification of inflationary models in the 
$n_{\rm S}$-$R$ plane in the high energy limit. 
The line $R=4(1-n_{\rm S})$
marks the border of large- and small-field models,
whereas the border of large-field and hybrid models
corresponds to $R=6(1-n_{\rm S})$.}
\label{brane}
\end{figure}

The exponential potential
\begin{equation}
V=V_{0}\exp \left(-\sqrt{\frac{16\pi}{\alpha}} 
\frac{\phi}{m_{{\rm Pl}}}\right)\,,
\label{exp}
\end{equation}
characterizes the border of large-field and hybrid models.
In the high-energy limit, we have 
$\epsilon=\eta=4\lambda/(V\alpha)$,
$n_{\rm S}-1=-16\lambda/(V \alpha)$ and 
$R=96\lambda/(V \alpha)$, thereby yielding
\begin{eqnarray}
R&=& 6 (1-n_{\rm S})\,.
\label{highcon}
\end{eqnarray}
In Fig.~\ref{brane} we plot the borders (\ref{nandR}) and 
(\ref{highcon}) together with the regions of three kinds 
of inflationary models.
The allowed range of hybrid models is wide
relative to large-field and small-field models.

\subsection{Large-field models}

Large-field models correspond to the parameter range with
$0<\eta \le \epsilon$.
The inflaton potential in these models is characterized as
\begin{equation}
V(\phi)=c\phi^p\,.
\label{polarge}
\end{equation}
For a fixed value of $p$ we have one free parameter, $c$, 
associated with the potential.

{}From Eqs.~(\ref{highnS}) and (\ref{highR}) one gets 
\begin{eqnarray}
\label{largenS}
n_{\rm S}-1 &=& -\frac{\lambda m_{{\rm Pl}}^2 p(2p+1)}
{2\pi c \phi^{p+2}}\,, \\
R &=& \frac{6\lambda m_{{\rm Pl}}^2 p^2}
{\pi c \phi^{p+2}}\,,
\end{eqnarray}
which means that $n_{\rm S}$ and $R$ are the functions of 
$\lambda$, $c$ and $\phi$.
{}From Eq.~(\ref{end}) we find that inflation ends at 
\begin{eqnarray}
\phi_f^{p+2}=\frac{5p^2M_5^6}{32\pi^2c}\,,
\end{eqnarray}
where we used the relation Eq.~(\ref{M5}).
The number of $e$-folds is 
\begin{eqnarray}
\label{efoldc}
N=\frac{4\pi c}{\lambda m_{{\rm Pl}}^2
p(p+2)}\phi^{p+2}-\frac{5p}{6(p+2)}\,.
\label{efold}
\end{eqnarray}
The second term ranges $5/12<p<5/6$ for 
$p>2$, thus negligible for $N \gtrsim 50$.
Then we have the following relation 
\begin{eqnarray}
\label{nRefold}
n_{\rm S}-1 &=& -\frac{2(2p+1)}{N(p+2)}\,, \\
R &=& \frac{24p}{N(p+2)}\,.
\label{nRefold2}
\end{eqnarray}
This is slightly different from what was obtained 
in Ref.~\cite{Liddle:2003gw} as 
we neglected the contribution coming from 
the second term in Eq.~(\ref{efold}).
For a fixed value of $p$, $n_{\rm S}$ and $R$
are only dependent on $N$.

\begin{figure}
\includegraphics[scale=0.6]{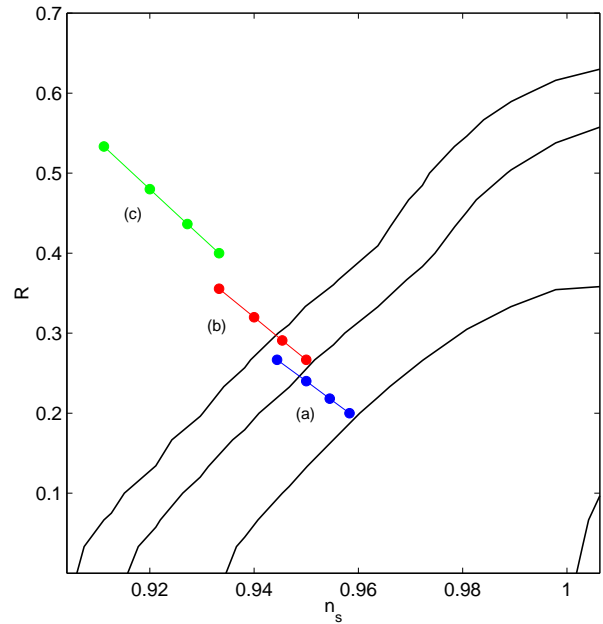} 
\caption{
Theoretical prediction of large-field models
together with the $1\sigma$, $2\sigma$ 
and $3\sigma$ observational contour bounds.
Each case corresponds to 
(a) $p=2$, (b) $p=4$ 
and (c) Exponential potential case ($p \to \infty$), 
respectively, with $e$-foldings
$N=45, 50, 55, 60$ (from top to bottom).
}
\label{nrlargefield}
\end{figure}

{}From Eqs.~(\ref{nRefold}) and (\ref{nRefold2}) we get 
\begin{eqnarray}
R=\frac{12p}{2p+1}(1-n_{\rm S})\,,
\label{line}
\end{eqnarray}
which corresponds to a straight line
for a fixed $p$.
For larger $p$, the tangent of the line Eq.~(\ref{line}) gets larger.
In Fig.~\ref{nrlargefield} we plot the values of $n_{\rm S}$
and $R$ for different values of $N$ and $p$.
Note that we consider several values of $e$-foldings which range
$45 \le N \le 60$, whereas in Ref.~\cite{Liddle:2003gw} this is 
fixed to be $N=55$.

When $p=2$ the theoretical predictions Eqs.~(\ref{nRefold}) 
and (\ref{nRefold2}) are within the $2\sigma$ observational
contour bound for $N>50$ as found from Fig.~\ref{nrlargefield}.
On the other hand the quartic potential ($p=4$)
is under strong observational pressure; the model is outside  
the $2\sigma$ bound for $N \lesssim 60$.
This case is disfavoured observationally, as in the case 
of standard inflation 
\cite{Peiris:2003ff,Barger:2003ym,Bridle,Kinney:2003uw,Leach:2003us,Tegmark}.

The exponential potential Eq.~(\ref{exp}) corresponds to the limit 
$p \to \infty$, in which case we have $n_{\rm S}-1=-4/N$ and
$R=24/N$ from Eqs.~(\ref{nRefold}) and (\ref{nRefold2}).
This case does not lie within the $2\sigma$ bound 
unless $N \gtrsim 90$.
Therefore the steep inflation \cite{Copeland:2000hn} 
driven by an exponential potential is excluded observationally.
Although inflation is realized even for $\alpha<1$ in Eq.~(\ref{exp})
in braneworld, the spectral index $n_{\rm S}$ and the ratio $R$
are shifted from the point $n_{\rm S}=1$ and $R=0$ due to the
steepness of the potential. 

For the potential Eq.~(\ref{polarge}) the amplitude of scalar 
perturbations is given as
\begin{eqnarray}
\label{scalas}
A_{{\rm S}} & \simeq &
\frac{64\pi^2c^2}{45pM_5^9}\phi^{2p+1} \, \\
& \simeq & 
\frac{64\pi^2c^2}{45pM_5^9}\left(\frac{3p(p+2)
M_5^6}{16\pi^2c}N\right)^{\frac{2p+1}{p+2}}\,,
\end{eqnarray}
from which we have
\begin{eqnarray}
\label{c}
c=\left(\frac{45pM_5^9A_{\rm S}}{64\pi^2}\right)^{\frac{p+2}{3}}
\left(\frac{3p(p+2)M_5^6N}{16\pi^2}\right)^{-\frac{2p+1}{3}}\,.
\end{eqnarray}
The COBE normalization corresponds to $A_{{\rm S}} \simeq 2\times
10^{-5}$ for $N \simeq 55$, which determines the amplitude $c$.

When $p=2$, the inflaton mass $m_\phi \equiv \sqrt{2c}$ 
is constrained from Eq.~(\ref{c}) as
\begin{eqnarray}
\label{mphi}
m_\phi \simeq 5 \times 10^{-5}M_5\,,
\end{eqnarray}
which is different from the case of standard inflation, $m_\phi \simeq 
10^{-6}m_{{\rm Pl}}$.
In the case of $p=4$ the self coupling, 
$\lambda_\phi \equiv 4c$, is constrained to be
\begin{eqnarray}
\label{lambda}
\lambda_\phi \simeq 8.2 \times 10^{-15}\,,
\end{eqnarray}
which is slightly smaller than the case of standard inflation, 
$\lambda_\phi \simeq 10^{-13}$.

As we have seen the model parameters can be strongly constrained
in large-field models.
This is due to the fact that we have only one free parameter, $c$, 
for the potential Eq.~(\ref{polarge}) and that both $n_{\rm S}$ and $R$
can be written by using the $e$-folding number only
even in the presence of the brane tension, $\lambda$.

\subsection{Small-field models}

Small-field models are characterized by the condition $\eta<0$,
which means that the second derivative of the potential is negative.
The potential in these models around the region $\phi=0$
is written in the form
\begin{eqnarray}
V\left(\phi\right) =V_0 \left[1 - \left({\phi
/ \mu}\right)^p\right]\,.
\label{posmall1}
\end{eqnarray}
A realistic model would consider a potential that has 
a local minimum, e.g.,
\begin{eqnarray}
V\left(\phi\right) =V_0 \left[1 - \frac12 \left(
\frac{\phi}{\mu} \right)^p\right]^2 \,,
\label{posmall2}
\end{eqnarray}
which is well approximated by the potential Eq.~(\ref{posmall1})
for $|(\phi/\mu)^p| \ll 1$.

New inflation is characterized by the potential Eq.~(\ref{posmall2}) 
with $p=2$. Natural inflation is slightly different from 
Eq.~(\ref{posmall2}), but it is approximated as 
Eq.~(\ref{posmall2}) by performing 
a Taylor expansion around $\phi=0$.
The potential of the tachyon field 
computed in the bosonic string field theory is 
given by \cite{Gerasimov:2000zp,Bento:2002np}
\begin{eqnarray}
V\left(\chi\right) =\tau_3 \left(\frac{\chi}{\chi_*}\right)^2
\left(1-2\,{\rm ln}\,\frac{\chi}{\chi_*} \right)\,,
\label{tachyon}
\end{eqnarray}
where $\tau_3$ is the D3 brane tension and $\chi_*=2l_{\rm s}
\sqrt{\tau_3}$ with $l_{\rm s}$ being a string-length scale.
This potential has a local maximum at $\chi=\chi_*$
and inflation is realized around this region
($\chi$ evolves toward the potential minimum at $\chi=0$).
We can perform a Taylor expansion around $\chi=\chi_*$,
which gives an approximate form of the potential
\begin{eqnarray}
V\left(\chi\right)  \simeq \tau_3 \left[1-2\left(\frac{\chi}{\chi_*}-1
\right)^2 \right]\,.
\label{tachyon2}
\end{eqnarray}
This reduces to the potential (\ref{posmall1}) with $p=2$
by rewriting $\tau_3=V_0$, $\phi=\chi-\chi_*$ 
and $\chi_*^2=2\mu^2$.
Therefore the tachyon potential Eq.~(\ref{tachyon}) belongs to small-field 
models
by shifting the potential maximum to $\phi=0$. 

Hereafter we shall consider the potential Eq.~(\ref{posmall2}) 
and assume the condition $|(\phi/\mu)^p| \ll 1$.
We obtain $n_{\Sc}$ and $R$ from 
Eqs.~(\ref{highnS}) and (\ref{highR}) as
\begin{eqnarray}
\label{nsmall}
n_{\Sc}&\simeq& 1-\frac{1}{2\pi}
\left(\frac{m_{{\rm Pl}}}{\mu}\right)^2
\frac{\lambda}{V_0}p \\ 
&\times& 
\biggl[(p-1)\left(\frac{\phi}{\mu}\right)^{p-2} 
+ \left(4p-\frac32 \right)
\left(\frac{\phi}{\mu}
\right)^{2(p-1)} \biggr]\,, \nonumber \\
R &\simeq& \frac{6}{\pi} 
\left(\frac{m_{{\rm Pl}}}{\mu}\right)^2
\frac{\lambda}{V_0}p^2
\left(\frac{\phi}{\mu}\right)^{2(p-1)}\,,
\label{Rsmall}
\end{eqnarray}
together with the amplitude of scalar perturbations
\begin{eqnarray}
A_{{\rm S}}^2 \simeq
\frac{64 \pi V_0^4 \mu^2}{75 m_{{\rm Pl}}^6
\lambda^3 p^2}
\left(\frac{\phi}{\mu}\right)^{2(1-p)}\,.
\label{amsmall} 
\end{eqnarray}

The field value $\phi_f$ takes a different form depending on 
the model parameters.
If the condition $r \equiv 5\lambda 
m_{{\rm Pl}}^2p^2/(24\pi \mu^2V_0) \ll 1$
is satisfied in Eq.~(\ref{end}), the end of inflation is 
characterized by $\phi_f/\mu \simeq 2^{-1/p}$.
This condition is automatically satisfied in the limit $\lambda/V_0 \to 0$.
In the case $r \gg 1$, which is possible for $\mu \ll m_{{\rm Pl}}$,
we approximately have 
$\phi_f/\mu \simeq [5\lambda m_{{\rm Pl}}^2p^2/(24\pi \mu^2V_0)]^{1/2(1-p)}$.
Hereafter we shall mainly discuss the case $r \ll 1$
and comment on the case $r \gg 1$ at the end.

\subsubsection{Case of $p=2$}

\begin{figure}
\includegraphics[scale=0.6]{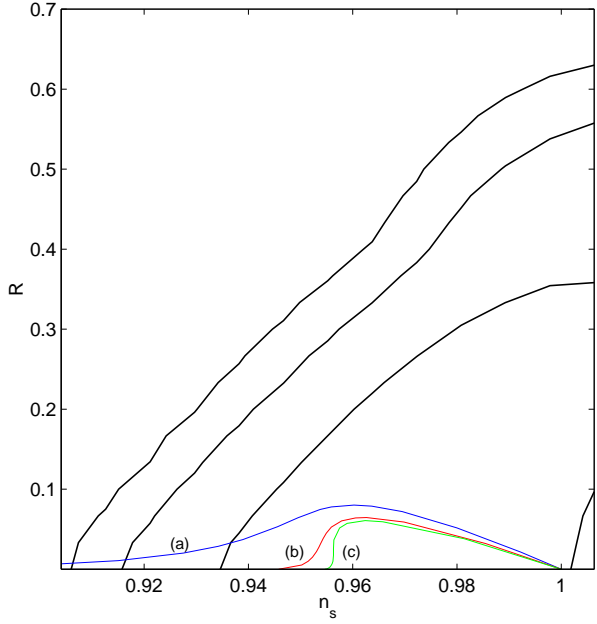} 
\caption{
Theoretical prediction of small-field models
together with the $1\sigma$, $2\sigma$ 
and $3\sigma$ observational contour bounds.
Each case corresponds
to (a) $p=2$, (b) $p=4$ and (c) $p=6$,
with $N=55$.}
\label{nRfig}
\end{figure}

When $p=2$ the field $\phi$ is expressed in terms of the $e$-folds $N$
from Eq.~(\ref{efolds})
\begin{eqnarray}
\phi \simeq \phi_f \exp \left(-\frac{\lambda m_{{\rm Pl}}^2}
{2\pi V_0\mu^2}N \right)\,.
\label{phifsmall1}
\end{eqnarray}
Making use of Eqs.~(\ref{nsmall}), (\ref{Rsmall}) and (\ref{amsmall})
with $\phi_f/\mu \simeq 2^{-1/p}$,
we get the following relations
\begin{eqnarray}
\label{Rsm}
R &=& \frac{128V_0^3}{25\lambda^2m_{{\rm Pl}}^4}
\frac{1}{A_{\rm S}^2}=12xe^{-Nx}\,, \\
n_{\rm S} &=& 1-\frac{13}{48}R-x\,,
\label{nsm}
\end{eqnarray}
where 
\begin{eqnarray}
\label{xdef}
x \equiv \frac{1}{2\pi}\left(\frac{m_{{\rm Pl}}}{\mu}\right)^2
\frac{\lambda}{V_0}p\,.
\end{eqnarray}
Then $R$ has a maximum value $R_{\rm max}=12/eN$ at 
$x=1/N$, which means that $R_{\rm max}$ is smaller than 
0.1 on cosmologically relevant scales 
($50 \lesssim N \lesssim 60$).
Comparing to the observational bounds shown in Fig.\,\ref{nRfig},
we find that the point with maximum $R$
is inside the $1\sigma$ curve.

Using the COBE normalized value 
$A_{\rm S} \simeq 2 \times 10^{-5}$
around $N=55$ with the condition $R \le 12/eN$, one gets 
\begin{eqnarray}
\left(\frac{V_0}{\lambda}\right)^2
\frac{V_0}{m_{{\rm Pl}}^4} \lesssim 
6.3 \times 10^{-12}\,. 
\label{Vzero}
\end{eqnarray}
This condition is violated in the high-energy limit $V_0/\lambda \to  
\infty$, which suggests that the information of the COBE normalization 
limits the strength $V_0/\lambda$.

In Fig.~\ref{nRfig} we show the two-dimensional plot of 
$n_{\rm S}$ and $R$
predicted by Eqs.~(\ref{Rsm}) and (\ref{nsm}) [see case (a)].
Compared to the 2D posterior observational constraints 
shown in Fig.\,\ref{nr}, theoretical predicted points 
are inside the $2\sigma$ curve as long as $|n_{\rm S}-1| \lesssim 0.09$.
This translates into the condition $x \lesssim 0.09$, i.e.,
\begin{eqnarray}
\left(\frac{m_{{\rm Pl}}}{\mu}\right)^2
\frac{\lambda}{V_0} \lesssim 0.28\,.
\label{muconst}
\end{eqnarray}
If we assume that the mass $\mu$ is smaller than 
$m_{{\rm Pl}}$, we get the constraints
$\lambda/V_0 \lesssim 0.28$ and 
$V_0/m_{{\rm Pl}}^4 \lesssim 5.0 \times 10^{-13}$
from Eqs.\,(\ref{Vzero}) and (\ref{muconst}).
In natural inflation the typical energy scale is 
the GUT scale, i.e., $V_0 \sim (10^{16}\,{\rm GeV})^4$,
corresponding to $V_0/m_{{\rm Pl}}^4 \sim 10^{-12}$.
The above upper limit of the energy scale in the brane case
is not much different from the one in standard inflationary
cosmology.

\subsubsection{Case of $p>2$}

When $p>2$ the terms in the square bracket of 
Eq.~(\ref{nsmall}) can be much smaller than unity 
for $|\phi/\mu| \ll 1$.
This means that the factor $x$ which appears in front of the square bracket of 
Eq.~(\ref{nsmall}) is not necessarily required to be  
smaller than unity in order to be compatible 
with observations. 

The field $\phi$ is written as 
\begin{eqnarray}
\left(\frac{\phi}{\mu}\right)^{2-p}=
\frac{\lambda m_{{\rm Pl}}^2p(p-2)N}{4\pi V_0\mu^2}
+\left(\frac{\phi_f}{\mu}\right)^{2-p}\,.
\label{phifsmall3}
\end{eqnarray}
Let us first consider the case of $x \ll 1$ and
$\phi_f/\mu \simeq 2^{-1/p}$.
Then we can express $R$ and $n_{\rm S}$
in terms of $x$:
\begin{eqnarray}
\label{Rsm2}
R &=& \frac{128V_0^3}{25\lambda^2m_{{\rm Pl}}^4}
\frac{1}{A_{\rm S}^2} \nonumber \\
&=& 12px \left( \frac{(p-2)Nx}{2}+2^{1-\frac{2}{p}}
\right)^{\frac{2(p-1)}{2-p}}\,, \\
n_{\rm S} &=& 1-\left(\frac13-\frac{1}{8p}\right)R
\nonumber \\
& & -\frac{2(p-1)x}{(p-2)Nx+2^{2(1-1/p)}}\,.
\label{nsm2}
\end{eqnarray}
Notice that $R$ is written in terms of $A_{\rm S}^2$
independent of the values of $p$.
The maximum value $R_{\rm max}$ gets gradually 
smaller with the increase of $p$.
When $p=4$, for example, one has $R_{\rm max}
\simeq 0.065$ for $N \simeq 55$, thereby yielding 
\begin{eqnarray}
\left(\frac{V_0}{\lambda}\right)^2
\frac{V_0}{m_{{\rm Pl}}^4} \lesssim 
5.1 \times 10^{-12}\,. 
\label{Vzero3}
\end{eqnarray}

Figure \ref{nRfig} indicates that the theoretical curves do not lie
in the region $n_{\rm S} \lesssim 0.94$ even with the increase in the value 
of $x$. Let us consider the case with $x \gg 1$
(corresponding to $r \gg 1$).
Since $\phi_f$ is estimated as  
$\phi_f/\mu \simeq [5\lambda m_{{\rm Pl}}^2p^2
/(24\pi \mu^2V_0)]^{1/2(1-p)}$ in this case, 
we have
\begin{eqnarray}
\label{Rsm3}
R &=& 
12px \left\{ \frac{(p-2)Nx}{2}+
\left(\frac{5px}{12}\right)^{\frac{2-p}{2(1-p)}}
\right\}^{\frac{2(p-1)}{2-p}}\!\!\!\!\!\!, \\
n_{\rm S}&=& 1-\left(\frac13-\frac{1}{8p}\right)R 
\nonumber \\
& & -\frac{2(p-1)x}{(p-2)Nx+2(5px/12)^{\frac{2-p}{2(1-p)}}}\,.
\label{nsm3}
\end{eqnarray}
In the limit $x \to \infty$, we find
$R \propto x^{p/(2-p)} \to 0$ and
\begin{eqnarray}
n_{\rm S} \to 1-\frac{2(p-1)}{(p-2)N}\,.
\label{nslimit}
\end{eqnarray}
When $N=55$ one has $n_{\rm S} \to 0.946$ for $p=4$
and $n_{\rm S} \to 0.955$ for $p=6$.
As seen in Fig.\,\ref{nRfig}, the curves
predicted by Eqs.~(\ref{Rsm2}) and (\ref{nsm2}) 
are inside the $1\sigma$ curve.
Therefore the model parameters are less constrained 
than in the case of $p=2$.
This is associated with the fact that the potential becomes
flat around $\phi=0$ for larger values of $p$, which 
does not exhibit strong deviation from $n_{\rm S}=1$
and $R=0$.

\subsection{Hybrid models}

Hybrid inflation is motivated by 
particle physics models which involve supersymmetry.
The potential of the original hybrid inflation proposed
by Linde is given as \cite{hybrid}
\begin{eqnarray}
V= \frac{\lambda_0}{4} \left(\chi^2-\frac{M^2}
{\lambda_0}\right)^2 +\frac12 g^2 \phi^2 \chi^2+
\frac12 m^2\phi^2\,.
\label{hybrid}
\end{eqnarray}
The supersymmetric scenario corresponds to 
$g^2=2\lambda_0$ \cite{hybrid2,Lyth:1998xn},
which is the case we shall consider hereafter.
Inflation occurs for $\phi>\phi_{f} \equiv M/g$, 
which is followed by the symmetry breaking driven by a
second scalar field, $\chi$. 
When the ``waterfall'' condition, 
$M^3 \ll \lambda_0 mm_{{\rm Pl}}^2$,
is satisfied, inflation soon comes to an end 
after the symmetry breaking \cite{hybrid}.  
This corresponds to the original version of the hybrid 
inflationary scenario where inflation ends due to the rapid rolling 
of the field $\chi$.
Setting $\chi \simeq 0$ in Eq.~(\ref{hybrid}), 
the effective potential for $\phi>\phi_{f}$ is written as
\begin{eqnarray}
V \simeq  \frac{M^4}{4\lambda_0}+\frac12
m^2\phi^2\,.
\label{hybrid2}
\end{eqnarray}

Notice that the second phase of inflation occurs 
after the symmetry breaking when the waterfall condition
is not satisfied. 
This corresponds to a double inflationary scenario
in which the second stage of inflation can 
affect the evolution of cosmological perturbations.
In fact, as shown in Ref.~\cite{Tsujikawa:2002qx},
the presence of the tachyonic instability for $\phi<\phi_f$
leads to the strong correlation between adiabatic and 
isocurvature perturbations, which can affect
the CMB power spectrum.

In this work we shall consider the case where perturbations
on cosmologically-relevant scales are generated before the symmetry 
breaking and also neglect the contribution of isocurvature perturbations.
Then the general hybrid inflation is given in the form
\begin{eqnarray}
V\left(\phi\right) = V_0
\left[1 + \left({\phi / \mu}\right)^p\right]\,,
\label{hybridgeneral}
\end{eqnarray}
which correspond to the parameter range of 
$\eta>\epsilon$ for $V/\lambda \gg 1$.
For the potential Eq.~(\ref{hybrid2}) we have 
three model parameters, $\lambda_0$, $M$ and $m$.
Note that $\phi_f$ is expressed in terms of $\lambda_0$ and $M$,
as $\phi_f=M/\sqrt{2\lambda_0}$.
This is equivalent to considering three parameters,
$V_0$, $\mu$ and $\phi_f$ for the potential Eq.~(\ref{hybridgeneral}).
Since we have one additional parameter compared to 
small-field models, it is expected that constraining 
the model is more difficult in this case. 

However one has
additional constraints on $V_0$ and $\mu$
in realistic supergravity models.
We can consider the following supergravity-motivated 
cases \cite{hybrid2,Linde:1997sj}
(corresponding to $p=2$ and $p=4$, respectively)
\begin{eqnarray}
\label{hybridone}
{\rm (i)}~~V(\phi) &=&V_{0} \left[1+8\pi
\left(\frac{\phi}{m_{{\rm Pl}}}\right)^2\right]\,, \\
{\rm (ii)}~~V(\phi)&=&V_{0} \left[1+
8\pi^2\left(\frac{\phi}{m_{{\rm Pl}}}\right)^4 \right]\,.
\label{hybridtwo}
\end{eqnarray}
In case (i) the second slow-roll parameter is given as 
$\eta=2/(1+V/2\lambda)$.
Then we have $\eta=2$ in the low-energy limit, 
which makes it difficult to achieve inflation
(the so-called $\eta$-problem).
This problem is overcome in the braneworld,
since $\eta \to 0$ for $V/\lambda \to \infty$.  
The case (ii) corresponds to inclusion of the supergravity
corrections to the effective potential in a globally-supersymmetric theory 
\cite{Linde:1997sj} 
(we neglected one-loop radiative corrections
calculated in Ref.~\cite{Linde:1997sj}).
We have $\eta=12\pi(\phi/m_{{\rm Pl}})^2/(1+V/2\lambda)$
in this case, which means that inflation is possible even 
in the low-energy limit as long as $\phi \ll m_{{\rm Pl}}$.

We shall first consider the general potential 
Eq.~(\ref{hybridgeneral}) without imposing 
the supergravity relations.
When $(\phi_f/\mu)^p$ is larger than unity in 
Eq.~(\ref{hybridgeneral}), this is not much 
different from large-field models discussed
in subsection A.
Therefore the condition $|(\phi/\mu)^p| \ll 1$
is assumed hereafter, in which case we have  
\begin{eqnarray}
\label{nhybrid}
n_{\Sc}&\simeq& 1+\frac{1}{2\pi}
\left(\frac{m_{{\rm Pl}}}{\mu}\right)^2
\frac{\lambda}{V_0}p \\ 
& & \times 
\biggl[(p-1)\left(\frac{\phi}{\mu}\right)^{p-2} 
+\left(2-5p\right)
\left(\frac{\phi}{\mu}
\right)^{2(p-1)} \biggr]\,,\nonumber 
\end{eqnarray}
where $R$ and $A_{\rm S}^2$ take the same forms as 
in Eqs.~(\ref{Rsmall}) and (\ref{amsmall}), 
thereby yielding 
\begin{eqnarray}
\label{Rhyb}
R = \frac{128V_0^3}{25\lambda^2m_{{\rm Pl}}^4}
\frac{1}{A_{\rm S}^2}\,.
\end{eqnarray}
The observational constraint imposes the condition 
$R \lesssim 0.57$ from Fig.~\ref{nr}, yielding
\begin{eqnarray}
\left(\frac{V_0}{\lambda}\right)^2
\frac{V_0}{m_{{\rm Pl}}^4} \lesssim 
4.5 \times 10^{-11}\,. 
\label{Vzero2}
\end{eqnarray}
This  is a general prediction of hybrid and small-field models.
Note that this value is more tightly constrained in small-field 
models as we showed in the previous subsection, 
but the hybrid models are somewhat different 
because of the additional model parameter.

Hereafter we consider the cases of $p=2$ and 
$p>2$ separately.

\subsubsection{Case of $p=2$}

In this case $\phi$, $R$ and $n_{\rm S}$ are expressed as
\begin{eqnarray}
\phi &=& \phi_f \exp \left(\frac{\lambda m_{{\rm Pl}}^2}
{2\pi V_0\mu^2}N \right)\,, \\
R &=& 24(\phi_f/\mu)^2xe^{Nx}\,, \\
n_{\rm S} &=& 1+x\left[1-8(\phi_f/\mu)^2e^{Nx}\right]\,,
\label{phfhyb1}
\end{eqnarray}
where $x$ is defined in Eq.~(\ref{xdef}).

\begin{figure}
\includegraphics[scale=0.6]{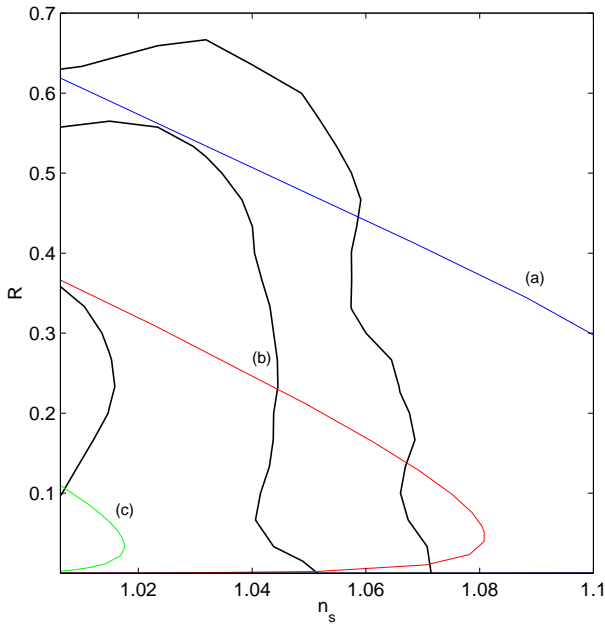} 
\caption{Theoretical prediction of hybrid models
together with the $1\sigma$, $2\sigma$ 
and $3\sigma$ observational contour bounds.
Each case corresponds to (a) $\phi_f/\mu=10^{-3}$, 
(b) $\phi_f/\mu=10^{-2}$ and (c) $\phi_f/\mu=10^{-1}$
with $N=55$.}
\label{nRhybrid1}
\end{figure}

In Fig.~\ref{nRhybrid1} we plot the above relations for 
several different values of $\phi_f/\mu$.
When $\phi_f/\mu \lesssim 10^{-3}$ the theoretical curve is
outside of the $2\sigma$ contour bound unless $x$ ranges 
$0 \le x \lesssim 0.05$
(note that $R$ is much smaller than 1 for  $0 \le x \lesssim 0.05$
in the case (a) of Fig.\,\ref{nRhybrid1}).
Therefore one gets the following constraint
\begin{eqnarray}
\left(\frac{m_{{\rm Pl}}}{\mu}\right)^2
\frac{\lambda}{V_0} \lesssim 0.16\,,~~
{\rm for}~~\phi_f/\mu \lesssim 10^{-3}\,.
\label{muconsthyb1}
\end{eqnarray}
The supergravity potential (\ref{hybridone}) corresponds to 
$\mu=m_{{\rm Pl}}/\sqrt{8\pi}$, which yields the 
constraint $\lambda/V_0 \lesssim 6.3 \times 10^{-3}$
from Eq.~(\ref{muconsthyb1}).
Making use of Eq.~(\ref{Vzero2}), we have a bound on the 
energy scale of inflation, $V_0/m_{{\rm Pl}}^4 \lesssim
1.6 \times 10^{-14}$. This is about $10^{-2}$ times lower
than the GUT scale, $V_0/m_{{\rm Pl}}^4 \sim 10^{-12}$.

When $\phi_f/\mu \gtrsim 10^{-3}$, Fig.\,\ref{nRhybrid1}
indicates that the theoretical curves begin to be within
the likelihood contour bounds.
In particular the curve is completely inside the $2\sigma$ 
bound for $\phi_f/\mu \gtrsim 0.03$, thus 
favoured observationally.
However we have one thing that must be considered with care.
There is a turnover for $n_{\rm S}$
coming from the second term in the square bracket 
of Eq.\,(\ref{nhybrid}).
Since we used the condition $(\phi_f/\mu)^2 \ll 1$
to derive this formula,  
Eq.\,(\ref{nhybrid}) is not valid when
$n_{\rm S}$ reaches close to 1 with the increase of $x$, 
corresponding to $x_{\rm M}=(2/N){\rm log} 
(\mu/2\sqrt{2}\phi_f)$.
The rough criterion for the validity of the approximation 
is $x \lesssim x_{\rm M}$, implying 
\begin{eqnarray}
\left(\frac{m_{{\rm Pl}}}{\mu}\right)^2
\frac{\lambda}{V_0} \lesssim 
\frac{2\pi}{N}\,{\rm log} \left(\frac{\mu}
{2\sqrt{2}\phi_f}\right)\,.
\label{muconsthyb2}
\end{eqnarray}
This comes from the requirement for hybrid inflation
$((\phi_f/\mu)^p \ll 1)$ so that the potential energy 
$V_0$ dominates in Eq.~(\ref{hybridgeneral}).
For the supergravity potential (\ref{hybridone})
with $N=55$, one gets $\lambda/V_0 \lesssim
4.5 \times 10^{-3} {\rm log}\,(\mu/(2\sqrt{2}\phi_f))$.
When $\phi_f/\mu \gtrsim 0.03$, this yields
$\lambda/V_0 \lesssim 1.2 \times 10^{-2}$.
Inflation is realized in the high-energy regime (corresponding 
to small $\lambda/V_0$) in this case, which gives the values of 
$n_{\rm S}$ and $R$ close to $n_{\rm S}=1$ and 
$R=0$.

\subsubsection{Case of $p>2$}

When $p>2$ we have
\begin{eqnarray}
\left(\frac{\phi}{\mu}\right)^{2-p} 
&=& \left(\frac{\phi_f}{\mu}\right)^{2-p}
-\frac{(p-2)Nx}{2} \,, \\
R &=& 12px \left[\left(\frac{\phi_f}{\mu}\right)^{2-p} 
-\frac{(p-2)Nx}{2}\right]^{\frac{2(p-1)}{2-p}}\!\!\!\!\!\!, \\
n_{\rm S} -1 &=& (p-1)x\left(\frac{R}{12px}\right)
^{\frac{p-2}{2(p-1)}}+\frac{2-5p}{12p}R\,.
\label{phifsmall4}
\end{eqnarray}
%

\begin{figure}
\includegraphics[scale=0.45]{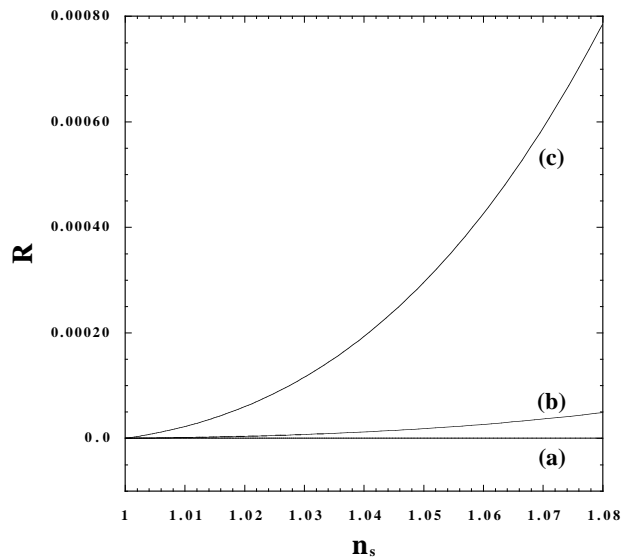} 
\caption{Theoretical prediction for $n_{\rm S}$ and $R$ in 
hybrid inflation models for $p=4$. 
Each case corresponds to (a) $\phi_f/\mu=10^{-2}$, 
(b) $\phi_f/\mu=5 \times 10^{-2}$ and 
(c) $\phi_f/\mu=10^{-1}$
with $N=55$.}
\label{nRhybrid2}
\end{figure}

Under the condition of $|\phi_f/\mu| \ll 1$, the ratio $R$ is much smaller 
than unity (see Fig.\,\ref{nRhybrid2}).
This means that we only need to consider the constraint on $n_{\rm S}$.
The spectral index is approximately written as
$n_{\rm S} -1 \simeq (p-1)x/[(\phi_f/\mu)^{2-p}-(p-2)Nx/2]$
for $|(\phi_f/\mu)^p| \ll 1$.
We have the $2\sigma$ constraint $n_{\rm S} \lesssim 1.05$ for 
$R \ll 1$ from Fig.\,\ref{nRhybrid1},
which leads to 
\begin{eqnarray}
x \lesssim \frac{0.05(\phi_f/\mu)^{2-p}}
{(p-1)+0.025(p-2)N}\,.
\label{xconhyb}
\end{eqnarray}
In the case of $p=4$ with $N=55$, this reduces to 
\begin{eqnarray}
\left(\frac{m_{{\rm Pl}}}{\mu}\right)^2
\frac{\lambda}{V_0} \lesssim 2.6 \times 10^{-2}
\left(\frac{\mu}{\phi_f}\right)^2\,.
\label{xconhyb2}
\end{eqnarray}
The supergravity potential Eq.~(\ref{hybridtwo}) corresponds to 
$\mu^2=m_{{\rm Pl}}^2/(\sqrt{8}\pi)$, in which case 
the condition Eq.~(\ref{xconhyb2}) is simplified as
\begin{eqnarray}
\frac{\lambda}{V_0} \lesssim 2.9 \times 10^{-3}
\left( \frac{\mu}{\phi_f} \right)^2\,.
\label{lamhyb}
\end{eqnarray}
Combining Eq.~(\ref{Vzero2}) with Eq.~(\ref{lamhyb}), we get 
the constraint 
\begin{eqnarray}
\frac{V_0}{m_{{\rm Pl}}^4} \lesssim
3.5 \times 10^{-16} \left(\frac{\mu}{\phi_f}\right)^4\,,
\label{hypo}
\end{eqnarray}
whose upper bound is dependent on the value $\phi_f/\mu$.

\section{Summary and Discussions}                            

In this paper we have investigated observational constraints 
on the Randall-Sundrum type II braneworld inflationary models.
The consistency relation Eq.~(\ref{consistency}) holds independent of
the strength of the brane tension $\lambda$, so that the same likelihood 
analysis can be employed as for standard inflationary models.
We carried out the likelihood analysis by varying the 
inflationary parameters as well as other cosmological parameters.
We take into account  several independent cosmological datasets
including the latest SDSS galaxy redshift survey \cite{SDSS}, and 
show that this leads to a tight constraint on the values of $n_{\rm S}$ 
and $R$ relative to including WMAP data alone \cite{Barger:2003ym}.
Our results are also consistent with the likelihood analysis
of the SDSS group \cite{Tegmark}. 
We also point out the importance of putting an appropriate prior
for the running of scalar perturbations, which otherwise leads to 
an unexpected shift toward larger likelihood values of $R$.

In braneworld the constraints on model parameters in terms of 
the underlying potentials are different compared to standard inflation.
We classified the models of inflation as large-field, small-field
and hybrid models, and constrained the model parameters
for each case.
In large-field models, which have only one free parameter in 
the potential, both $n_{\rm S}$ and $R$ are the function 
in terms of the $e$-folds $N$ only.
This simple property allows us to place strong constraints
on large-field models. While the quadratic potential is 
within the observational $2\sigma$ bound for $N \gtrsim 50$,
the quartic potential is disfavoured since the predicted curve is 
outside the $2\sigma$ bound for $N<60$
(see Fig.\,\ref{nrlargefield}).
The steep inflation driven by an exponential potential 
is far outside the $3\sigma$ bound, thus excluded observationally.

In small-field models one has an additional parameter associated
with the potential, which implies 
that it is more difficult to constrain model parameters 
compared to large-field models.
In spite of this, we can place an upper bound on the energy 
scale of inflation using the information of $A_{\rm S}^2$
and $R$, see Eqs.~(\ref{Vzero}) and (\ref{Vzero3}).
The $p=2$ case for the potential Eq.~(\ref{posmall2}) is not excluded 
as long as the condition Eq.~(\ref{muconst}) is satisfied.
The $p>2$ case does not possess additional restrictions
on model parameters, since the theoretical prediction is 
within the $1\sigma$ contour bound (see Fig.\,\ref{nRfig}).

Hybrid models are more involved due to the fact that 
there is one more additional parameter associated with 
the end of inflation ($\phi_f$).
Nevertheless we placed limits on the energy scale of inflation
and also obtained the relationship between $\mu, \lambda/V_0$
and $\phi_f$ from observational bounds in the 
$n_{\rm S}$--$R$ plane
[see Eqs.~(\ref{Vzero2}), (\ref{muconsthyb2}) and 
(\ref{xconhyb2})].
These relations can be simplified in supergravity models
due to an additional relation between $V_0$ and $\mu$.

Although the constraint on each inflation model in braneworld
differs from the one in standard inflationary cosmology,
the likelihood values of inflationary parameters 
($A_{\rm S}$, $R$, $n_{\rm S}$, $n_{\rm T}$,
$\alpha_{\rm S}$, $\alpha_{\rm T}$) are the same in both cases.
In order to pick up the signature of braneworld, it is required
to break through the degeneracy of the consistency relation.
This degeneracy is associated with the fact that 5-dimensional 
observables smoothly approach the 4-dimensional counterpart
in an exact de-Sitter embedding, 
as we decouple the brane from the bulk with an 
increasing brane tension.
However it was recently shown in Ref.~\cite{Seery}
that this does not hold for a marginally-perturbed de-Sitter
geometry and the relationship between observables
is dependent on the brane tension
(see also Ref.~\cite{Calcagni}).
This can provide one possible way to distinguish 
between braneworld and standard inflation
from different constraints on observables.

While we concentrated on single-field inflationary 
scenarios in this work, the CMB power spectrum 
is generally modified if isocurvature perturbations dominate 
adiabatic ones \cite{Langlois}.
In the low-energy case the correlation between adiabatic 
and isocurvature perturbations is strong for the double inflation 
model with potential given by Eq.~(\ref{hybrid}) \cite{Tsujikawa:2002qx}.
When the second stage of inflation occurs after the symmetry 
breaking, it is important to follow the dynamics of curvature
perturbations ${\cal R}$ precisely, since ${\cal R}$
is no longer conserved in the context of multi-field 
inflation \cite{multi}. 
The enhancement of curvature perturbations reduces the 
relative amplitude of tensor to scalar perturbations,
which leads to the modified consistency 
relation \cite{Bartolo}
\begin{eqnarray}
R=-8n_{\rm T}(1-r_C^2)\,,
\label{mconsistency}
\end{eqnarray}
where $r_C$ is the correlation between adiabatic and isocurvature
perturbations. Although it is not obvious whether the same consistency 
relation holds or not in braneworld, especially when the second scalar field 
corresponds to the brane modulus, it would be interesting to find out 
the signature of braneworld in such generic cases. 
See Ref.~\cite{Ashcroft} for recent work in this direction.

In this paper we have only considered braneworld effects on generating the 
initial power spectra, while in general there may be 5D effects at late times 
impacting on, for example, CMB anisotropies.
Recently there have been several attempts to give a quantitative prediction 
of the CMB power spectrum by solving a bulk geometry using a
low-energy approximation \cite{Kanno} in a two-brane system
\cite{Koyama:2003be,Rhodes:2003ev}.
While this approach involves some unresolved issues such as 
the stabilization of the modulus (radion), 
this is the first important step to understand the effect of 
the 5D perturbations. In particular it was shown in Ref.~\cite{Koyama:2003be}
that the effect of the Weyl anisotropic stress leads to the modification of 
the CMB temperature anisotropy around the first doppler peak,
while the perturbations on larger scales are not altered.
It would be certainly of interest to extend this analysis to the 
high-energy regime in order to fully pick up the effect of extra dimensions
on CMB anisotropies.
We hope that this will open up a possibility to distinguish the braneworld
scenario from other inflationary models motivated by, e.g., 
quantum gravity \cite{QG} or noncommutative 
geometry \cite{noncominf}.

\acknowledgments

We are indebted to Sam Leach for substantial help in implementation 
of the Monte 
Carlo Markov Chain analysis used in this paper. 
Antony Lewis and David Parkinson 
also provided kind support
in implementing and interpreting the likelihood analysis.
S.T. thanks Roy Maartens, Takahiro Tanaka, and 
David Wands for useful discussions, and acknowledges
financial support from JSPS (No.\,04942). 
S.T. is also grateful to all members in IUCAA for their warm 
hospitality and especially to Rita Sinha for her kind support
in numerics. 

\end{document}